# Magneto-optical study of magnetization reversal asymmetry in exchange bias

A. Tillmanns, S. Oertker*, B. Beschoten**, and G. Güntherodt

II. Physikalisches Institut, RWTH Aachen, 52056 Aachen, Germany

C. Leighton***, and Ivan K. Schuller

Department of Physics, University of California - San Diego, La Jolla, CA 2093-0319

J. Nogués

Institució Catalana de Recerca i Estudis Avançats (ICREA) and Departament de Física,

Universitat Autònoma de Barcelona, 08193 Bellaterra, Spain

The asymmetric magnetization reversal in exchange biased Fe/MnF$_2$ involves coherent (Stoner-Wohlfarth) magnetization rotation into an intermediate, stable state perpendicular to the applied field. We provide here experimentally tested analytical conditions for the unambiguous observation of both longitudinal and transverse magnetization components using the magneto-optical Kerr effect. This provides a fast and powerful probe of coherent magnetization reversal as well as its chirality. Surprisingly, the sign and asymmetry of the transverse magnetization component of Fe/MnF$_2$ change with the angle between cooling and measurement fields.

A ferromagnetic (FM) layer in contact with an antiferromagnetic (AFM) one experiences a shift of the hysteresis loop along the field axis, due to the so-called exchange bias (EB) [1,2]. One of the intriguing features of EB is a pronounced asymmetry in the hysteresis loops, a very unusual phenomenon in magnetism since all other magnetic materials exhibit symmetric reversal. The most prominent example of an asymmetric reversal occurs in Fe/MnF$_2$, which exhibits a pronounced step on only one side of the hysteresis loop [3]. Polarized neutron reflectometry (PNR) showed that this step is related to coherent rotation of the magnetization [3,4], which is pinned in a potential minimum transverse to the applied magnetic field. As PNR is not sensitive to the direction of the transverse moment it cannot determine the chirality of the magnetization vector upon magnetization reversal.

Although asymmetric magnetization reversal has been claimed from PNR [4], viscosity [5], and anisotropic magnetoresistance [6] measurements, the obvious experiment using magneto-optical Kerr effect (MOKE) has not yet been performed. MOKE studies, accessing both longitudinal and transverse magnetization components are scarce for exchange biased systems [7-9]. It is therefore important to analyze how coherent magnetization rotation manifests itself in a well-defined MOKE signal and whether the absence of such a signature therefore implies the existence of domain wall nucleation and propagation processes.

Here, we use MOKE in separate longitudinal and transverse geometries to study magnetization reversal in the model exchange-bias system Fe/MnF$_2$(110). This fast technique allows us to determine the chirality of the magnetization reversal via the *sign* of the transverse magnetization component. We derive analytical conditions to unambiguously identify the Kerr signature of pure longitudinal ($M_L$) or transverse ($M_T$) magnetization components. These conditions allow for the determination of the orientation and relative magnitude of the in-plane magnetization components at all fields during magnetization reversal. Surprisingly, the hysteresis loop asymmetry critically depends on the angle $\varphi_H$ between the in-plane measurement field and the cooling field direction. Within a few degrees $M_T$ changes its sign. For 90° it appears on both sides of the loop with the same sign of rotation, contrary to a Stoner-Wohlfarth (360°) reversal process.

Polycrystalline Fe on epitaxial, twinned, $MnF_2$(110) has been grown in the structure MgO(100)/$ZnF_2$(110)/$MnF_2$(110)/Fe/Al(cap layer), with thicknesses of ⋯/25/65/12/3 nm, respectively. Details of the sample preparation and structural characterization are given elsewhere [3].

MOKE measurements were performed with an in-plane magnetic field oriented at 45° with respect to the [001] direction of the AFM twins. This is crucial to observe the asymmetry in the hysteresis loop [3]. Kerr loops were taken in two separate configurations (Fig. 1(a)). In the longitudinal geometry (I) the scattering plane is parallel to $M_L$, while in the transverse geometry (II) it is parallel to $M_T$. For both geometries the linearly polarized incident light can be continuously rotated from $s$ to $p$ polarization using a $\lambda/2$ retarding plate. Kerr rotation of the reflected laser beams can be simultaneously detected in both geometries. For detection (Fig. 1(b)), the beam reflected from the sample passes a Glan-Thompson polarizing beam splitter where it is separated into two orthogonal polarized beams which are focused by lenses onto diodes A and B of a diode bridge. The light intensities at the diodes, $I_A$ and $I_B$, and the difference signal $I_{A-B}$ are simultaneously measured using a lock-in amplifier. Prior to each measurement the diode bridge is balanced ($I_{A-B} = 0$) using a $\lambda/2$ retarding plate. All Kerr loops were taken at $T = 20$ K after field cooling (FC) through the Néel temperature ($T_N = 78$ K) in $H_{FC} = 1$ kOe in the film plane at 45° to the twins aligned along the [001] direction of $MnF_2$(110).

Fig. 1(c) shows Kerr rotation ($I_{A-B}$) data using s-polarized incident light for both MOKE configurations (polarization angle $\Psi = 90°$ with respect to each scattering plane). For the longitudinal configuration we observe a strongly asymmetric hysteresis loop with a distinct plateau on the negative field side of the loop. The loop shape is identical to that taken with a SQUID magnetometer under identical conditions (not shown), indicating that this Kerr loop is a measure of $M_L$. In the transverse configuration a pronounced peak is observed only in the field range where the longitudinal signal shows the horizontal plateau. Elsewhere the Kerr rotation is zero. Such a peak was observed earlier, on an identical sample, by PNR, and was shown to originate from $M_T$. Thus, MOKE in the transverse configuration probes unambiguously the transverse magnetization component $M_T$ due to coherent magnetization rotation.

Fig. 1(d) shows a series of Kerr loops (all taken in the longitudinal configuration) as a function of the incident polarization direction from *s*- to *p*-polarized state ($\psi = 90°$ to $0°$). The observation of identically shaped loops but with reversed sign for pure s- and p-polarized light is an intriguing feature. Even more surprisingly, the loop at $\psi = 45°$ is very similar to the one in the transverse configuration (Fig. 1(c)). This indicates that both $M_L$ and $M_T$ can actually be measured in a single geometrical setup simply by changing $\psi$. For all other $\psi$, the Kerr loops appear to be a superposition of the loops taken at $\psi = 0°$ and $45°$.

More detailed results are summarized in Fig. 2 in both longitudinal (left panel) and transverse (right panel) geometries, for $\psi = 90°$ (s-pol., Fig. 2(a,d)), $45°$ (Fig. 2(b,e)), and $0°$ (p-pol., Fig. 2(c,f)). The curves for the intensities $I_A$ and $I_B$ measured in diodes *A* and *B* (Fig. 1(b)), and the difference signal $I_{A-B}$, are vertically shifted. In the transverse configuration the Kerr loops ($I_{A-B}$) are identical for *s*- and *p*-polarization except for their sign. In addition, the loop at $\psi = 45°$ has the shape found for *s*-polarization in the longitudinal configuration.

Most previous analytical MOKE descriptions deal with Kerr rotation detection with a crossed analyzer [10-13], as opposed to the diode bridge technique used in our work. Here, we derive analytical expressions for observations of the pure components $M_L$ and $M_T$, using the diode bridge detection method.

In the *longitudinal geometry* the optical path is given by [11]

$$\frac{\vec{E}}{E_0} = \begin{bmatrix} \cos\alpha & -\sin\alpha \\ \sin\alpha & \cos\alpha \end{bmatrix} \begin{bmatrix} 1+M_T T & M_L L \\ -M_L L & S \end{bmatrix} \begin{bmatrix} \cos\psi \\ \sin\psi \end{bmatrix} \quad (1)$$

with $S = r_{ss}/r_{pp}$, $T = \Delta r/r_{pp}$, $L = r_{sp}^l/r_{pp}$, where $r_{pp}$ and $r_{ss}$ are the isotropic Fresnel reflection coefficients for ($\psi = 0°$) p- and ($\psi = 90°$) s-polarized incident light and $\Delta r$ and $r_{sp}^l$ are the transverse and longitudinal magneto-optic coefficients, respectively. $\psi$ is the polarization angle of the incident laser beam (as defined above) and $\alpha$ is the angle of polarization rotation, which is required to balance the diode bridge. For $\psi = 0°$ $I^L/I_0 = |E/E_0|^2$ at the diodes is given by

$$\frac{I_{A,p}^L}{I_0} = M_T \operatorname{Re}(T) - M_L \operatorname{Re}(L) + \ldots,$$

$$\frac{I_{B,p}^L}{I_0} = M_T \operatorname{Re}(T) + M_L \operatorname{Re}(L) + \ldots.$$

Higher order terms in $M$ have been neglected. Interestingly, the difference $I_{A-B,p}^L$ is a measure of $M_L$ only, because

$$\frac{I_{A-B,p}^L}{I_0} = -2M_L \operatorname{Re}(L), \quad (2a)$$

consistent with Fig. 2(c). For other polarizations

$$\frac{I_{A-B,45°}^L}{I_0} = M_T \operatorname{Re}(T) + M_L \operatorname{Re}(L)(1+\operatorname{Re}(S)) + M_L \operatorname{Im}(L)\operatorname{Im}(S), \text{ and} \quad (2b)$$

$$\frac{I_{A-B,s}^L}{I_0} = -2M_L \operatorname{Re}(L)\operatorname{Re}(S) - 2M_L \operatorname{Im}(L)\operatorname{Im}(S). \quad (2c)$$

If $I_{A-B,s}^L = -I_{A-B,p}^L$ (*condition 1*) is satisfied (Fig. 1(d)) the intensities become

$$\frac{I_{A-B,45°}^L}{I_0} = M_T \operatorname{Re}(T), \text{ and} \quad (3a)$$

$$\frac{I^L_{A-B,s}}{I_0} = 2M_L \,\mathrm{Re}(L). \tag{3b}$$

Hence, for $\psi = 45°$, $I^L_{A-B,45°}$ (Fig. 2(b)) is a pure measure of $M_T$.

In the *transverse geometry* at $\psi = 45°$, we find that $I^T_{A-B,45°}$ is a pure measure of $M_L$ if $I^T_{A-B,s} = -I^T_{A-B,p}$ (*condition 2*) which is also satisfied experimentally (Fig. 2 (e)).

We point out that the pure $M_T$ signal in the longitudinal geometry, or the pure $M_L$ signal in the transverse geometry, will appear at $\psi \neq 45°$ if $I^{L,T}_{A-B,s} \neq -I^{L,T}_{A-B,p}$. Although the determination of this angle is beyond the scope of this paper, we do stress the fact that incorrect conclusions may be obtained if "conditions 1 and 2" are not met. Note, that this measurement technique is not quantitative vector MOKE, because the ratio of the transverse and longitudinal magnetization components depends on the ratio of the magneto-optic coefficients Re(L) and Re(T) in Eqs. (2a), (3a) and (3b) which are not known. The ratio of these coefficients, however, does in first order not depend on the magnetic field H and therefore the measurements as a function of H are independent of the above deficiency.

Most surprisingly, varying the in-plane applied magnetic field by $\varphi_H = \pm 3°$ with respect to the cooling field direction the transverse magnetization component changes sign, i.e. its chirality reverses (Fig.3). The fact that for $\varphi_H = \pm 3°$ no $M_T$ signal appears on the right side of the hysteresis loop does, however, not imply that the magnetization reversal proceeds via domain wall nucleation and propagation [14].

At an applied field direction of, e.g., $\varphi_H = 90°$ the transverse component appears on both sides of the loop with the same sign, i.e. rotating within the same half plane ($M_T (H) < 0$) in contrast to the full 360° rotation of a Stoner-Wohlfarth reversal. Similar observations were found for Co/CoO and Fe/FeF$_2$ showing that this behavior is not particular to the system discussed here. This will be the subject of a further publication [15].

The types of reversal modes observed for finite $\varphi_H$ in twinned MnF$_2$(110)/Fe confirm qualitatively the theoretical predictions for an untwinned EB system [16]. In this model the absence of a transverse component ($M_T = 0$) on one side of the loop is attributed to a nonuniform reversal mode.

In conclusion, systematic MOKE investigations on the model exchange bias system Fe/MnF$_2$(110) provide unambiguous experimental and analytical identification of well-defined pure transverse and longitudinal magnetization components. Direct comparison with an analytical model yields straightforward conditions that allow for the unambiguous decomposition of the magnetization into its longitudinal and transverse components. We have demonstrated that the step in the hysteresis loop of Fe/MnF$_2$(110) is related to a pure transverse magnetization component. A surprising sign change of the transverse component with small angular variations between the cooling and measuring fields is observed. For 90° the transverse magnetization component appears on both sides of the loop with the same sign, unlike a Stoner-Wohlfarth-type magnetization reversal process. These observations confirm theoretically predicted reversal modes as function of the angle between the cooling and measurement fields.

Work supported by the: DFG/SPP1133, European Community's Human Potential Program/NEXBIAS, Spanish CICYT, Catalan DGR, and US-DOE. I. K. Schuller thanks the von Humboldt Foundation for support and RWTH faculty and researchers for hospitality during a sabbatical stay in Aachen.

*present address: Institut für Biologische Informationsverarbeitung IBI, Forschungszentrum Jülich, 52425 Jülich, Germany

**Email address: bernd.beschoten@physik.rwth-aachen.de

***present address: Department of Chemical Engineering and Materials Science, University of Minnesota

FIGURE Captions

FIGURE 1

(a) MOKE setup in longitudinal (I) and transverse (II) geometries. (b) Schematic diagram of the diode bridge detector for MOKE ($\lambda/2$ = half-wave plate, G-T = Glan-Thompson prism, which separates orthogonal linear polarizations). The light reflected from the sample enters from the lower left side. MOKE loops of Fe/MnF$_2$ at $T = 20$ K for (c) *s*-polarized incident light in the longitudinal (I) and the transverse (II) configuration and (d) different polarization directions of the incident beam in the longitudinal (I) configuration.

FIGURE 2

MOKE loops of Fe/MnF$_2$ (110) at $T = 20$ K in the longitudinal (left panel) and transverse (right panel) configuration for different incident polarization directions. Both diode signals $I_A$ and $I_B$ and their difference signal $I_{A-B}$ are shifted and shown for each polarization.

FIGURE 3

Sensitivity of transverse MOKE signal of Fe/MnF$_2$(110) at $T = 20$ K to variations of the angle between the directions of the cooling and measurement fields.

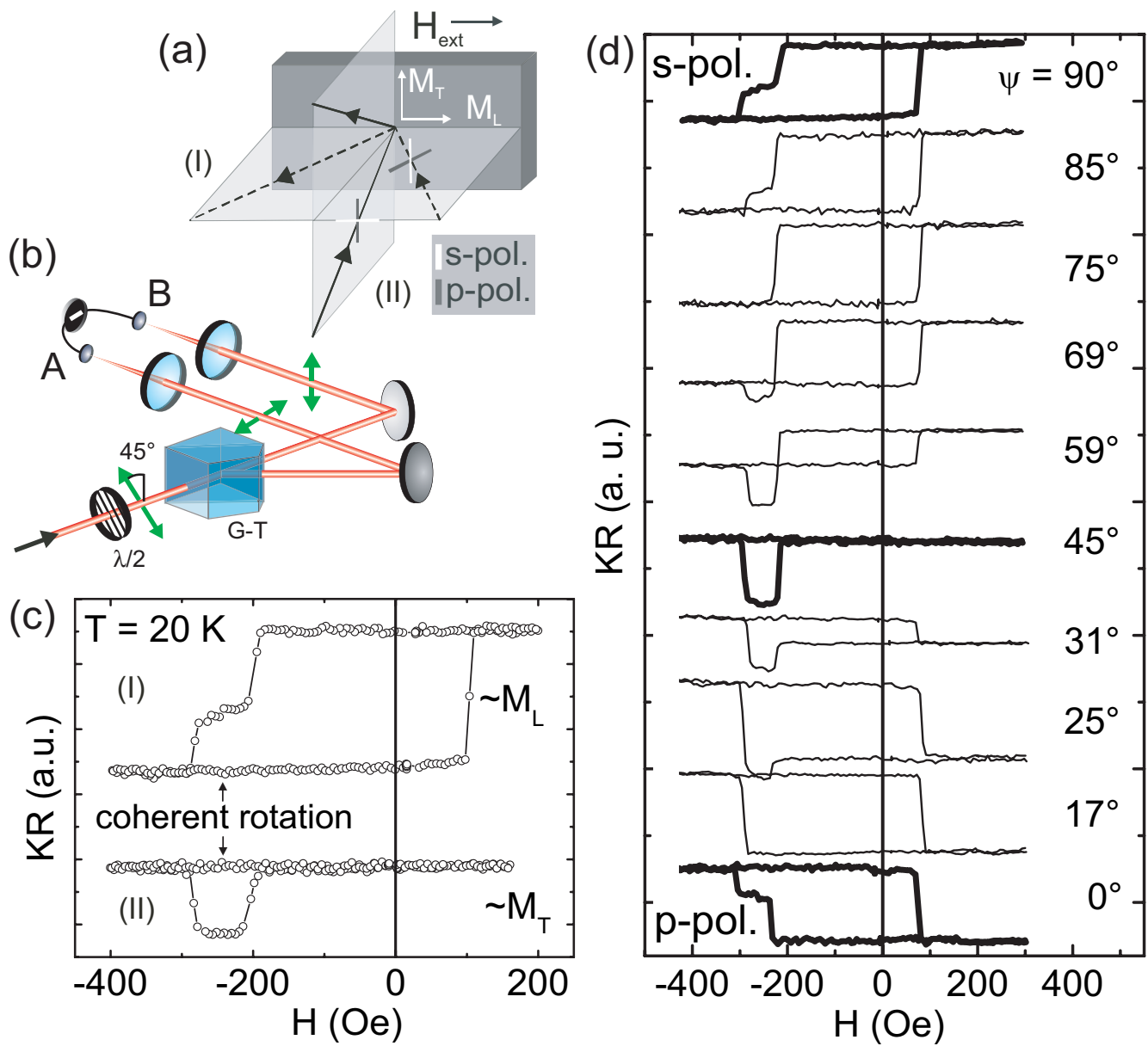

Figure 1: A. Tillmanns et al.

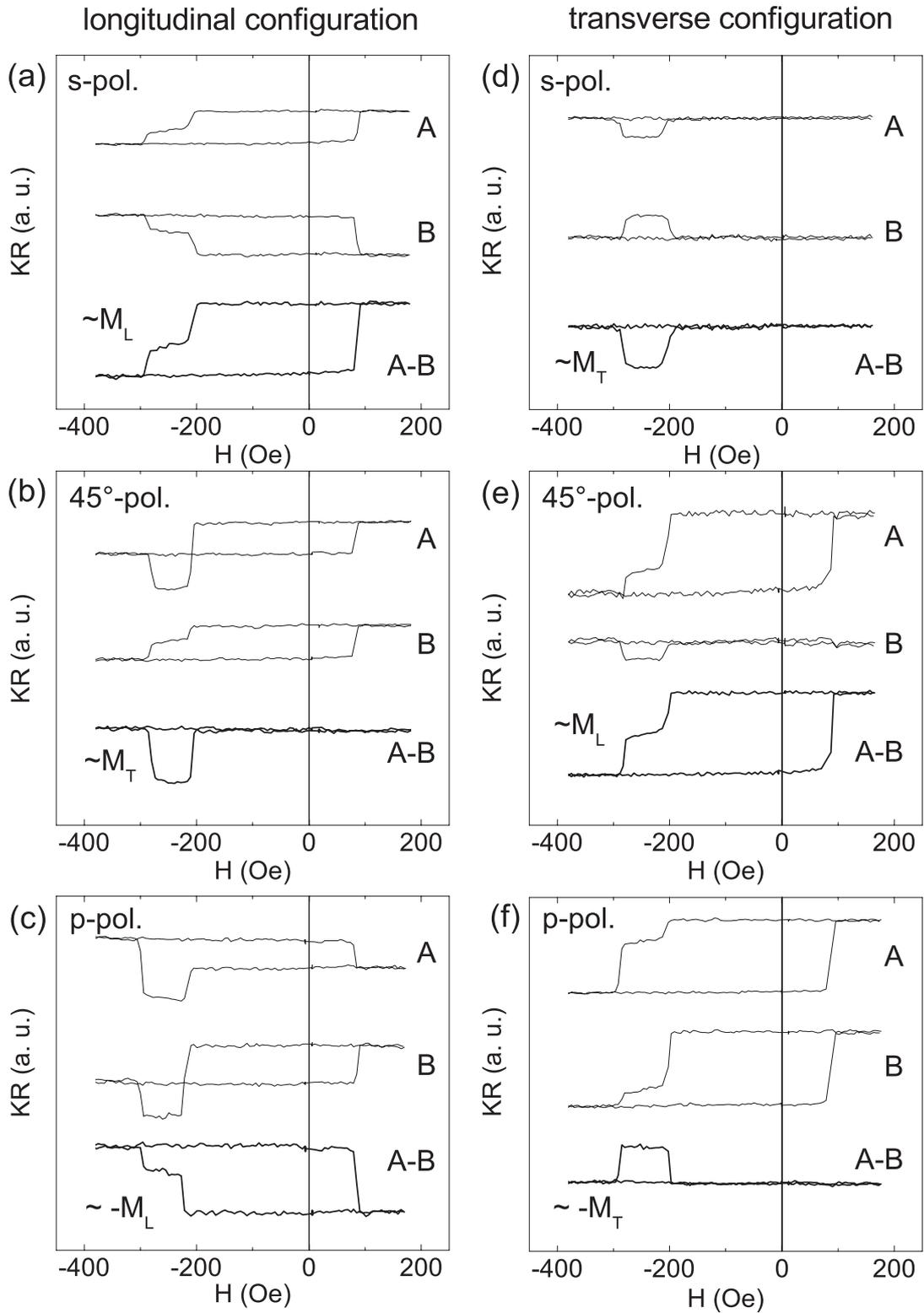

Figure 2: A. Tillmanns et al.

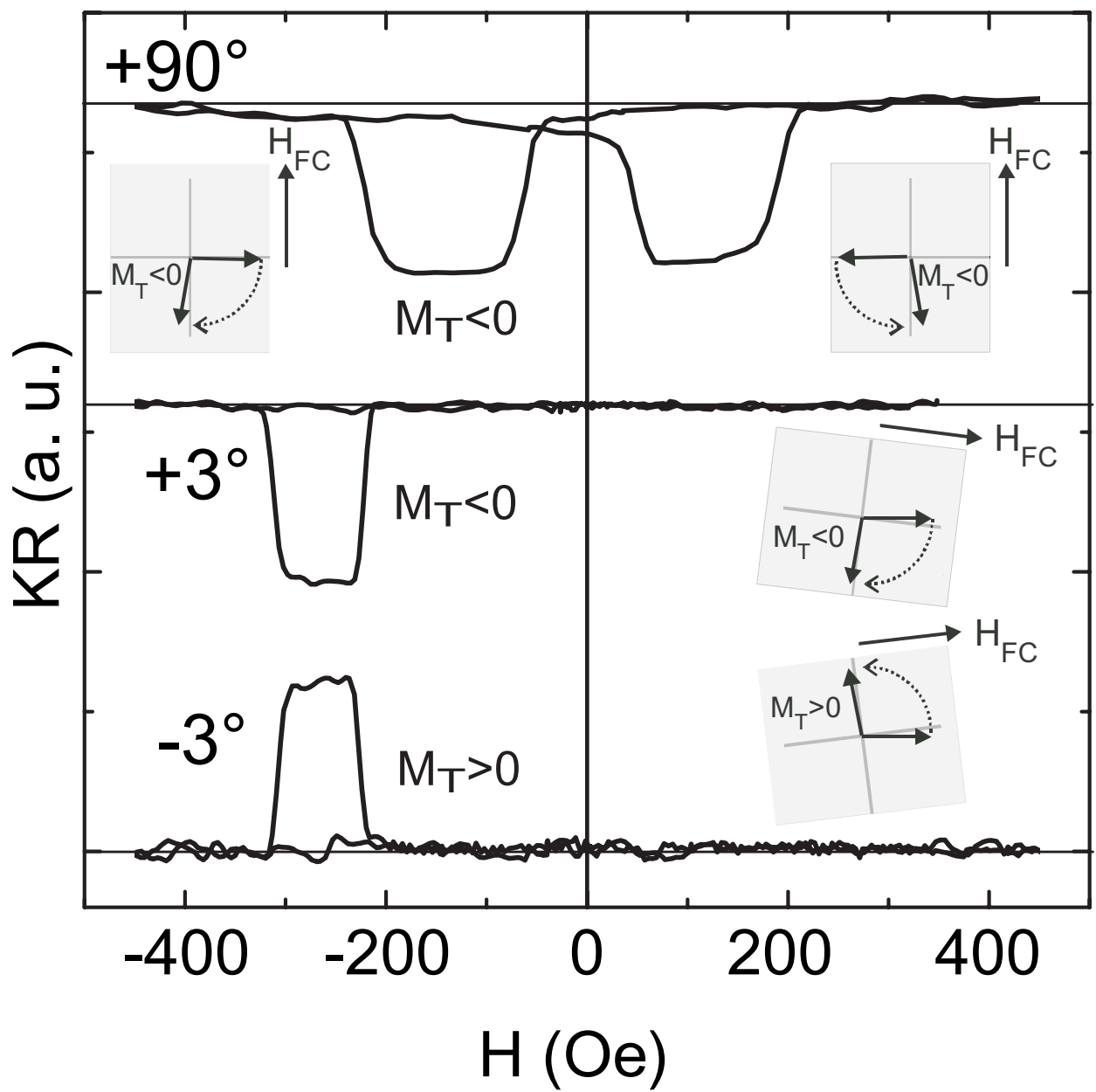

Figure 3: A. Tillmanns et al.